\documentclass[12pt,a4paper]{article}
\usepackage{amsmath,amsfonts,amssymb,bm}
\usepackage{epsfig}

\def\lsi{\raise0.3ex\hbox{$<$\kern-0.75em\raise-1.1ex\hbox{$\sim$}}}
\def\gsi{\raise0.3ex\hbox{$>$\kern-0.75em\raise-1.1ex\hbox{$\sim$}}}
\newcommand{\lsim}{\mathop{\lsi}}
\newcommand{\gsim}{\mathop{\gsi}}

\def\Pom{{\bf I\!P}}

\begin{document}

\begin{center}

{\Large \bf From soft to hard regime \\ in elastic
pion-pion scattering \\ above resonances}

\vspace{0.6cm}

{\large
A. Szczurek $^{1,2}$ }

\vspace{0.2cm}

$^{1}${\em Institute of Nuclear Physics\\
PL-31-342 Cracow, Poland\\}
$^{2}${\em Rzesz\'ow University\\
PL-35-959 Rzesz\'ow, Poland\\}

\vspace{0.6cm}

{\large
N.N. Nikolaev $^{3,4}$ and J. Speth $^{3}$ }

$^{3}${\em Institut f\"ur Kernphysik (Theorie), Forschungszentrum
J\"ulich,\\
D-52425 J\"ulich, Germany \\}
$^{4}${\em L.D. Landau Institute for Theoretical Physics \\
Chernogolovka, Moscow Region 142 432, Russia \\}

\end{center}


\begin{abstract}
We discuss the onset of the dominance of the
Glauber-Gribov-Landshoff (GGL) component of pQCD hard two-gluon
(2G) exchange contribution to hard elastic $\pi\pi$ scattering at
moderate energies . Such a hard  $\pi\pi$ scattering could via
final state interaction on the $\gamma \gamma \rightarrow \pi^+
\pi^-$ reaction in which pQCD quark-exchange contribution is known
to be short of strength. While in the nonrelativistic
approximation the GGL amplitude is known to be free of suppression
by the pion form factor, we show that in the relativistic
light-cone approach it acquires a residual, albeit a weak,
suppression. Furthermore,z the same mechanism it is free of the
end-point contributions. We evaluation the GGL amplitude with a
model light-cone wave function consistent with the pion charge
form factor data. The soft contribution to elastic $\pi\pi$
scattering is estimated based on the  $NN$ and $\pi N$ total cross
section data and Regge factorization, which gives the $\pi\pi$
total cross sections consistent with the ones deduced earlier from
the absorption model analysis of the $\pi N \to X N, X \Delta$
data. We evaluate the large-$|t|$ tail of the soft amplitude
within the Regge absorption models. We find that while in the same
sign  $\pi^\pm\pi^\pm$ scattering the hard GGL mechanism takes
over at $|t|\gsim$ 3 GeV$^{-2}$, in the opposite-sign
$\pi^\pm\pi^\mp$ scattering the hard GGL mechanism $|t|\lsim$ 4
GeV$^{-2}$.
\end{abstract}

\section{Introduction}

The pion-pion scattering, although not directly accessible experimentally,
is of special theoretical interest. At low energy it is the fundamental
testing ground of chiral perturbation theory
\cite{Weinberg,ChPT,effective_field_theories}.
For the extension into the resonance region one includes
explicit resonance fields in conjunction with suitable unitarization
models \cite{unitarization} or invokes  meson-exchange interactions
\cite{meson_exchange_models} tested in low and intermediate
energy $NN$ and $\pi N$ interactions. Soft, i.e., small angle, pion-pion
scattering at moderate energies above the prominent resonances falls
into the domain of the Regge theory. In the soft region the scattering
amplitudes fall rapidly with $|t|$ because of the form factor effects.
Large-angle, hard, scattering will eventually be dominated by pQCD
mechanisms.

At low and resonance energies one of important sources on $\pi\pi$
scattering is the extraction of the pion exchange contribution
to $\pi + p \rightarrow \pi + \pi + N$ (see e.g.\cite{Brookhaven} and
references therein). There is only very limited information on
$\pi \pi$ scattering above resonances. Here the information
about the $\pi \pi$ total cross section comes from the absorption
model analysis of the experimental data on $\pi N \to X N, X \Delta$
reactions \cite{ZS84}.

There are no direct experimental data on hard $\pi\pi$ scattering
which is of special interest within pQCD. At low $|t|$ the
$t$-dependence of the scattering amplitude is controlled by the
size of the beam and target hadrons. For instance, within the
Glauber-Gribov multiple scattering theory of composite objects,
the impulse approximation amplitude is proportional to the product
of one-body form factors of the beam and target
\cite{Glauber,Gribov}. Within the same multiple scattering theory,
elastic scattering of the $n$-body beam on the $n$-body target
receives the special contribution form the n-fold rescattering in
which different constituents of the beam scatter off different
constituents of the target. This special contribution does not
depend on the size of the beam and target, i.e. is free of the
form factor suppression. In the realm of pQCD such a three-gluon
exchange mechanism of $pp,\bar{p}p$ scattering has been discussed
by Landshoff \cite{Landshoff}. In scattering of the two-body pions
the related Glauber-Gribov-Landshoff (GGL) mechanism emerges
already at the level of the two-gluon exchange, i.e., in the Born
term for the pQCD hard pomeron exchange. For this reason, one can
expect a precocious dominance of the GGL mechanism in hard $\pi
\pi$ scattering.

In this communication we discuss the onset of the dominance of the
GGL mechanism in $\pi\pi$ scattering at $\sqrt{s} \lsim$ 10 GeV.
Such an analysis requires an understanding of the large-$|t|$ tail
of soft elastic scattering, which we evaluate within the Regge
absorption models. The Glauber-Gribov arguments for the absence of
the form factor suppression of the GGL amplitude were implicitly
based on the nonrelativistic (NR) approximation for composite
systems. We find that in contrast to NR approximation, in the
light-cone approach there is a residual, albeit weak, form factor
suppression of the GGL amplitude. Although the direct experimental
study of hard $\pi\pi$ elastic scattering is not feasible, this
process can contribute via final state interaction to hard charge
exchange reaction $\gamma \gamma \rightarrow \pi^+ \pi^-$ which
has been observed experimentally \cite{gamma_gamma} with the cross
section much larger than perturbative QCD quark-exchange
predictions \cite{pQCD}. New, preliminary data from DELPHI
\cite{Grzelak} suggest even stronger excess with respect to pQCD.
Our interest in hard $\pi\pi$ scattering has been motivated by
this missing strength of pQCD predictions, because  soft
$\gamma\gamma \to \pi\pi$ process followed by hard $\pi\pi$
elastic scattering could account for the missing strength.

The presentation of the paper is organized as follows. We start
with the evaluation of the pQCD 2G exchange amplitude in section
2. In section 3, assuming Regge factorization, we derive
parameters of soft $\pi \pi$ scattering, based on experimental
data for $\pi N$ and $N N$ scattering. The role of multiple soft
and hard rescatterings on angular distributions of pions is
discussed in section 4.

\section{The pQCD two-gluon exchange}

\begin{figure}

\mbox{
\epsfsize 2.0in
\epsfbox{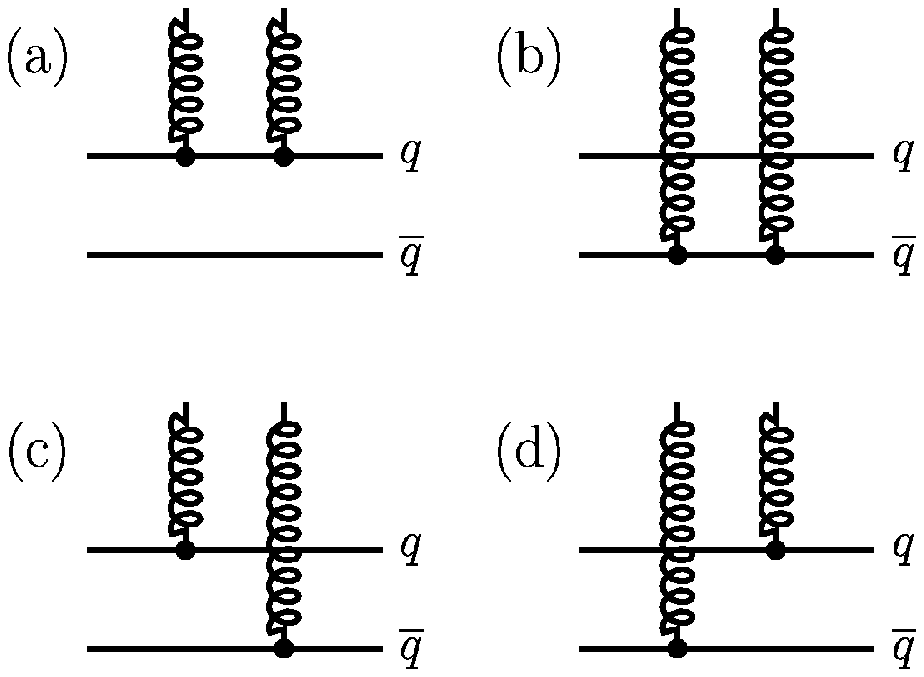}
}

\caption{\it The pQCD diagrams for the pion impact factors.}
\label{fig_gluon}
\end{figure}

Let us start with evaluation of the pQCD two-gluon contribution to
the elastic pion-pion scattering. We treat the pion as the
quark-antiquark state. The relevant pQCD diagrams which contribute
to the pion impact factor are shown in Fig.\ref{fig_gluon}. A
calculation of Born amplitudes in the non-relativistic
approximation for the target and beam hadrons can be found
elsewhere \cite{2Gluon}. In the present communication we use both
nonrelativistic and the light-cone description of the pion.

Making use of the Sudakov technique \cite{QED}, one readily obtains
the impact factor representation of the pion-pion scattering
amplitude
\begin{eqnarray}
A(\vec{q}) = i s \cdot \frac{2}{9} \cdot \frac{1}{(2\pi)^2} \cdot
\int d^2 \kappa \; g_s^2(\kappa_1^2) g_s^2(\kappa_2^2) \Phi_{\pi
\rightarrow \pi}^{2G}(\vec{q},\vec{\kappa}) \Phi_{\pi \rightarrow
\pi}^{2G}(\vec{q},\vec{\kappa})
\nonumber \\
\frac{1}{(\vec{q}/2 + \vec{\kappa})^2}
\frac{1}{(\vec{q}/2 - \vec{\kappa})^2} \;,
\label{amplitude_impactfactors}
\end{eqnarray}
where $\vec{\kappa}_{1/2} = \frac{\vec{q}}{2} \pm \vec{\kappa}$ are the
exchanged-gluon momenta, which are purely transverse,
2/9 is the QCD color factor for the $\pi\pi$ scattering process considered
and $g_s$ is the QCD strong charge\footnote{In our practical
calculations the QCD coupling constant is frozen in the infra-red region.}.
 Please note that the coupling
constants $g_s$ have been taken out from the impact factors.
We use the standard normalization of amplitudes such that
\begin{equation}
\frac{d\sigma}{dt} = \frac{1}{16 \pi s^2} |A(t,u,s)|^2 \; .
\label{amp_convention}
\end{equation}
 We approximate the pion by its $q\bar{q}$ Fock state. A
relatively strightforward calculation of the diagrams which define
two-gluon pionic impact factor gives \cite{Schwiete}
\begin{eqnarray}
\Phi_{\pi \rightarrow \pi}^{2G}(\vec{q},\vec{\kappa}) &=&
 \int_0^1 \frac{dz d^2 k}{z^2 (1-z)^2}
\nonumber \\
& &\left\{ \Psi(z, \vec{k}) \Psi(z, \vec{k}+z \vec{q})
\cdot [m_Q^2 + \vec{k} \cdot (\vec{k}+z \vec{q})] \right.
\nonumber \\
&+& \Psi(z, \vec{k}) \Psi(z, \vec{k}-(1-z) \vec{q})
\cdot [m_Q^2 + \vec{k} \cdot ( \vec{k}-(1-z) \vec{q})]
\nonumber \\
&-& \Psi(z, \vec{k}) \Psi(z, \vec{k}+(z-\tfrac{1}{2}) \vec{q}
-\vec{\kappa})
\cdot [m_Q^2 + \vec{k} \cdot (\vec{k}+(z-\tfrac{1}{2}) \vec{q}
-\vec{\kappa})]
\nonumber \\
&-& \left. \Psi(z, \vec{k}) \Psi(z, \vec{k}+(z-\tfrac{1}{2}) \vec{q}
+\vec{\kappa})
\cdot [m_Q^2 + \vec{k} \cdot (\vec{k}+(z-\tfrac{1}{2}) \vec{q}
+\vec{\kappa})]
\right\}  \; .
\nonumber \\
\label{LC_impact_factor}
\end{eqnarray}

Here the pion-quark-antiquark vertex is taken of the form
$i\Gamma_{\pi}(M^2)\overline{\Psi}\gamma_{5}\Psi$. In terms of the
quark \& antiquark helicities $\lambda$ the $\pi q({\bf
k})\bar{q}(-{\bf k})$ vertex has the form (\cite{NSS01}, for the related
discussion see Jaus \cite{Jaus})
\begin{equation}
\overline{q}_{\lambda}({\bf k})\gamma_{5}q_{\bar{\lambda}}(-{\bf
k}) = \frac{\lambda}{ \sqrt{z(1-z)}} [m_{Q} \delta_{\lambda
-\bar{\lambda}} - k(-\lambda) \delta_{\lambda \bar{\lambda}}]\, ,
\label{eq:2.2}
\end{equation}
where $m_{Q}$ is the quark mass, taken equal for the up and down
quarks, $k(\lambda)=\sqrt{2}{\bf k}\cdot {\bf \epsilon}_{\lambda}$
and $\bf \epsilon_{\lambda}=\frac{1}{\sqrt{2}} (\lambda {\bf
\epsilon}_{x}+i{\bf \epsilon}_{y})$ is the familiar polarization
vector for the state of helicity $\lambda$. In transitions of
spin-zero pions into $q\bar{q}$ states with the sum of helicities
$\lambda+\bar{\lambda}=\pm 1$ the latter is compensated by the
orbital momentum of quark and antiquark.

The radial wave function of the pion in the momentum space is
defined in terms of the $\pi q\bar{q}$ vertex function as
\begin{equation}
\psi_{\pi}(z,{\bf k})={N_{c}\Gamma_{\pi}(R,M^{2}) \over 4\pi^3
z(1-z)( M^2 -m_{\pi}^2)} \label{eq:2.10}
\end{equation}
where $M$ is the invariant mass of the constituent quark-antiquark
system,
$$
M^2={\vec{k}^2 +m_{Q}^2 \over z(1-z)}\, , $$ $\vec{k}$ is the
transverse momentum of the quark, $z$ and $(1-z)$ are fractions of
the lightcone momentum of the pion carried by a quark and
antiquark, respectively.

The first two terms in (\ref{LC_impact_factor}) corresponding to
the impulse approximation (IA) diagrams (a) and (b) in Fig.1
are equal to the one-body pion form factor,
\begin{equation}
G_{\pi}(\vec{q}^2 ) = \int_0^1 \frac{dz}{z^{2}(1-z)^{2}} \int d^2
k [m_Q^2 + \vec{k} \cdot (\vec{k}+z \vec{q})] \Psi(z,\vec{k})
\Psi(z,\vec{k} + (z \vec{q}) \; .
 \label{one_body_formfactor}
\end{equation}
Such contributions, when exchanged gluons couple to one constitutent,
are 
typical for additive quark models and
are suppressed at large transverse momentum $\vec{q}$
by the form factor.

The last two terms in (\ref{LC_impact_factor}) correspond to the
diagrams (c) and (d) in Fig.1 and in conjunction with their
counterpart in the target impact factor give the GGL contribution
to elastic scattering amplitude. Now notice that in the NR
approximation $z={1\over 2}$, so that the contribution of the
diagrams (c) and (d) would not depend on $\vec{q}$ altogether.
Then a comparison with the one-body form factor
(\ref{one_body_formfactor}) evaluated in the same approximation
would give the impact factor of a simple form
\begin{equation}
\Phi_{\pi \rightarrow \pi}^{2G}(\vec{q},\vec{\kappa}) = 2 [
G_{\pi}(q^2) - G_{\pi}(4 \kappa^2) ] \; , \label{NR_impact_factor}
\end{equation}
where the factor 2 in front is the number of constituents in the pion.

Beyond the NR approximation,  the terms $(z-{1\over 2})\vec{q}$ in
the arguments of the last two terms in
Eq.~(\ref{LC_impact_factor}) are substantial at large $\vec{q}$
and make them decreasing with $\vec{q}$. Furthermore, at very
large $q$ the dominant contribution to the GGL amplitude would
come from $(z-{1\over 2})\lsim {1\over r_{\pi}q}$, where $R_{\pi}$
is the pion radius, so that the GGL amplitude is free of
contributions from the end points $z\to 0$ and $z\to 1$ (for the
discussion of the end-point properties of the GGL contribution to
the $pp$ scattering see \cite{Botts}).

For the calculation of pQCD 2G-exchange in the NR case it is
sufficient to know the form factor of the pion. We take here
simply the $\rho$-dominance monopole parameterization. In the
light-cone approach, the Ansatz for the WF must be subjected to
the normalization condition,
\begin{equation}
\int_{0}^{1} \frac{dz}{z(1-z)} \int d^2 k \; M^2
|\Psi(z,\vec{k})|^2 = 1 \; , \label{normalization_condition}
\end{equation}
and its parameters, i.e., the effective quark mass $m_Q$ and the
pion radius $R_{\pi}$ must be constrained by the $\pi \to \mu \nu$
decay constant (we use the PDG convention $F_{\pi}=131$ MeV
\cite{PDG})
\begin{equation}
F_{\pi}= \int d^2{\bf k} dz m_{f} \psi_{\pi}(z,{\bf k})\, ,
\label{eq:2.11}
\end{equation}
the charge radius of the pion as defined by
eq.~(\ref{one_body_formfactor}), the slope of the form factor of
the $\pi^0 \to \gamma\gamma^*$ transition and the decay width
$\Gamma_{\pi^0 \rightarrow \gamma \gamma}$ \cite{ Schwiete,Jaus,NSS01}. 
The results for these low-energy
parameters for the pion from the WF parameterization \cite{NSS01}
with $m_Q$ = 0.215 GeV and $R_{\pi}$ = 2.2 GeV$^{-1}$ are cited in
table 1. This WF provides a satisfactory description of the charge
form factor of the pion into the semihard region of $q^2$, as seen
from fig~2, in which we compare the prediction ftom eq.~(4) to the
recent experimental data from the Jefferson Laboratory [20].

The above discussion of the 2G exchange holds in the pQCD domain
of $|t| \gg 1$ GeV$^2$. In order to extrapolate the hard pQCD
scattering contribution to the forward scattering amplitude and to
the total cross section, one needs to impose an infrared (IR)
regularization to mimic the finite propagation radius of color
fields. We model this by the ``Debye'' screening in the gluon
propagator,
$$
{1\over \kappa^2 } \Longrightarrow {1\over \kappa^2 +m_{g}^2}\,,
$$
where $m_{g}=R_{c}^{-1}$. It was found in \cite{NZZ} that the pQCD
2G exchange with the Debye screening radius $R_{c}\approx 0.27$ fm
provides a viable boundary condition for the color dipole BFKL
description of the proton structure function. More recently, very
close values of the screening radius $R_{c}$ were found from an
analysis \cite{Meggiolaro} of color field correlations in lattice
QCD, (see also a recent review \cite{Field} on the effective mass
of the gluon).
The different analyses lead to $m_g$ of about 0.75 GeV. The band
between the results for $m_g$ = 0.5 GeV and $m_g$ = 1.0 GeV can be
regarded as a conservative estimate for uncertainties of the
two-gluon exchange amplitude at subasymptotic $|t|$.\bigskip\\

\begin{table}[h]
\caption{Low energy parameters of the pion from the Ansatz
\protect{\cite{NSS01}} for the pion WF.}
\begin{center}
\begin{tabular}{|c|c|c|}
\hline
observable & experiment & WF Ansatz \protect{\cite{NSS01}} \\
\hline
$<r_{\pi}^2> [fm^2]$ &  0.451 $\pm$ 0.05 &  0.442 \\
$f_{\pi}$  [MeV]     &  130.7 $\pm$ 0.15 &  121   \\
$\Gamma_{\pi^0 \rightarrow \gamma \gamma}$ [eV] & 7.8 $\pm$ 0.6 &  7.67 \\
$\Lambda_{\pi^0}$ [MeV] &  750 $\pm$ 30 & 640 \\
\hline
\end{tabular}
\end{center}
\end{table}


%

The numerical results from pQCD 2G exchange for $\pi\pi$
scattering are shown in Fig.\ref{fig_mg} for different values of
the IR screening parameter $m_g$, for the nonrelativistic (left
panel) and light-cone (right panel) approach.
The hatched area gives an idea of the uncertainties due to the not
precisely known correlation radius. For a comparison, we show also
the soft cross section evaluated in the Regge impulse
approximation, i.e., the single pomeron and reggeon exchange.
 At $|t| \gtrsim$ 2 GeV$^2$ the hard contribution takes over the soft
contribution evaluated in the impulse approximation, for the
discussion of how the soft-hard interplay is affected by
absorption correction see in more detail below.

The anatomy of the pQCD 2G exchange is shown in more detail in
fig.~4. Here the dashed line shows the result found if only the IA
components of  result obtained if only the impulse approximation
diagrams (a) and (b) of fig.~1 are kept in the impact factor of
both pions. The dotted line shows the pure  GGL contribution, when
only the diagrams (c) and (d) of fig.~1 are kept in the impact
factor of both pions.  The contribution from the impulse
approximation components of the impact factors dominates at small
to moderate values of $|t| \lsim $ 0.5 GeV$^2$, where the
nonrelativistic and light-cone amplitudes are nearly identical.
The slight variation from NR to LC cases is due to the fact that
the one-body form factor given by the LC Ansatz decreases somewhat
faster than the $\rho$-pole formula. The GGL mechanism starts
taking over at $|t| \gsim $ 1.0 GeV$^2$. Here a comparison of the
NR and LC cases shows clearly a substantial suppression of the GGL
contribution by the $q$-$z$ correlations inherent to the LC case.
One should note, however, that destructive interference of the
impulse approximation and GGL components of the impact factor
remains noticeable even at large $|t|$ $\sim$ 4 GeV$^2$.
\section{Soft Regge-pole amplitudes}

The understanding of the onset of hard pQCD regime requires
evaluation of the large-$|t|$ tail from soft non-perturbative
interactions. In the nonperturbative region of small transferred
momenta in baryon-baryon and meson-baryon elastic scattering one
is bound to phenomenological parameterizations such as the
seasoned absorption Regge model \cite{Regge, Collins}.
The powerful Regge factorization enables one to estimate
the $\pi \pi$ scattering amplitudes from the experimental data
on $\pi N$ and $NN$ scattering.
The absorption corrections model the large-$|t|$ tail of
the scattering amplitude.

In the case of pion-pion scattering in the considered region of
energies the soft pomeron exchange must be supplemented
by the subleading isoscalar ($f$) and isovector ($\rho$) reggeon
exchanges. For the purposes of our analysis we resort to
the simplest Regge-inspired phenomenological form:
\begin{eqnarray}
A_{\Pom}(t) &=& i \; C_{\Pom} \cdot (s/s_0)^{\alpha_{\Pom}(t)}
\cdot F_{\Pom}^2(t) \; ,
\nonumber \\
A_{f}(t)  &=& -\eta_f(t) \; C_f \cdot (s/s_0)^{\alpha_f(t)}
\cdot F_f^2(t) \; ,
\nonumber \\
A_{\rho}(t) &=& -\eta_{\rho}(t) \; C_{\rho} \cdot (s/s_0)^{\alpha_{\rho}(t)}
\cdot F_{\rho}^2(t) \; ,
\label{reggeon_amplitudes}
\end{eqnarray}
where $\eta_f$ and $\eta_{\rho}$ are somewhat simplified signature factors:
\begin{equation}
\eta_R = \exp(i \phi_R(t)) \; ,
\end{equation}
with
\begin{equation}
  \phi_R(t) =
  \begin{cases}
    &- \frac{\pi}{2} \alpha_R(t) \qquad \text{ ~~for positive
      signature} \; , \\
    &- \frac{\pi}{2}\left[ \alpha_R(t)-1 \right] \text{ for negative
      signature}
        \; .
  \end{cases}
\end{equation}
The pion-reggeon(pomeron) vertex form factor $F_i$ in
(\ref{reggeon_amplitudes}) is not calculable from the first
principles and one is bound to parameterizations driven by the
educated guess and plausible restrictions on the large-$|t|$
behaviour of the form factors. One the one hand, one would like
soft amplitudes to decrease at large-$|t|$ faster than the
perturbative ones, and one customarily uses the exponential
parametrization $F(t) =\ exp(\frac{B}{4}t)$. On the other hand, the
differential cross section of the elastic scattering exhibits at
small $|t|$ a curvature which is somewhat better described if one
would take the inverse power parameterization $F(t) =1/(1-B_{mon})$. 
Such a monopole form factor for the
pion-reggeon vertex should not be extended indiscriminately to
large-$|t|$, though, otherwise soft and hard form factors would
have had unwanted similar asymptotic behaviour. In what follows,
we shall evaluate the absorption corrections for the exponential
soft pion-reggeon(pomeron) form factor.

In the Regge-pole exchange approximation, the powerful Regge
factorization allows to relate the residues $C_i$ of the Regge
pole contributions to different scattering processes.  are related
by the Regge factorization. In our case of $\pi \pi$ scattering
residues at $t=0$ can be evaluated from those for $\pi N$ and $NN$
scattering as:
\begin{equation}
C^{\pi\pi}_i = {(C^{\pi N}_i)^2 \over  C^{NN}_i}
\label{factorization}
\end{equation}
for each reggeon considered $i = \Pom, f, \rho$. Although
absorption corrections, i.e., the Regge cut contributions, violate
the exact Regge factorization, it still remains a viable
approximation for small $t$, as  well documented in many reactions
\cite{Regge,Collins}. Specifically, although the absorption
corrections to the Regge-pole approximation can be as large as
10-20 \%, they are believed not to change strongly from one
reaction to another \cite{Regge,Collins}. For our purpose of
evaluating the soft background to hard pQCD pion-pion scattering,
we are content with the 10-20 \% accuracy. Then, based on the
known Regge phenomenology of $\pi N$ and $NN$ scattering
\cite{DL92}, from (\ref{factorization}) we find for $\pi\pi$
scattering $C_{\Pom}$ = 8.56 mb, $C_f$ = 13.39 mb and $C_{\rho}$ =
16.38 mb. We take for the pomeron trajectory $\alpha_{\Pom}(0)$ =
1 and $\alpha_{\Pom}^1$ = 0.25 GeV$^{-2}$ and for both subleading
trajectories $\alpha_R(0)$ = 0.5 and $\alpha_R^1$ = 0.9
GeV$^{-2}$, i.e. values well known from the Regge phenomenology
\cite{Collins}. In the above evaluation we can safely neglect have
a small, see fig.~3, pQCD 2G exchange contribution to the $\pi N$
and $NN$ total cross section (see also \cite{NSZpion}).
For the diffraction slope, the same Regge factorization entails
\begin{equation}
B_{\pi\pi} \approx 2B_{\pi N} -B_{NN} \, , \label{SLOPES}
\end{equation}
which suggests $B_{\pi\pi} \sim (6-8)$ GeV$^{-2}$. The diffraction
slopes are fairly sensitive to the absorption corrections, though,
see the discussion below. For the purposes of the present
exploratory analysis, it is sufficient to a universal slope for all
the pion-reggeon(pomeron) vertex form factors,
$B_{\Pom}=B_{f}=B_{\rho}=B$. This slope is the main adjustable
parameter of our soft amplitudes.

The total single-reggeon exchange amplitude is now
\begin{equation}
A_{soft}^{1-st}(t) = A_{\Pom}(t) + A_f(t) + \xi A_{\rho}(t)  \; ,
\end{equation}
where $\xi$ = -1 for $\pi^+ \pi^-$, $\xi$ = 0 for $\pi^{\pm} \pi^0$
and $\xi$ = 1 for two identical pions.

We note in passing that QCD motivated
models for soft pomeron exchange were discussed
in \cite{NSZpion,QCD_nonperturbative}.
To a crude approximation, such models respect the quark additivity
and in their extension to the $\pi\pi$ scattering are similar
to the Regge factorization approach.

The role of the soft Regge amplitude at small $t$ is illustrated
in  fig.~3, where the dashed line shows the soft Regge
contribution to the differential cross section evaluated with the
exponential form factor $F$ and B = 6 GeV$^{-2}$ , as suggested by
(\ref{SLOPES}) \footnote{Here we did not include yet absorption
corrections, which allowance for which the preferred slope of the
Regge-pole amplitudes is rather close to B$\approx$ 4 GeV$^{-2}$.}
As anticipated the soft-reggeon exchanges dominate at small $|t|$
over the discussed in the previous section two-gluon exchange. The
situation reverses at larger $|t|$, but as we shall see below, the
exact pattern of the soft-to-hard transition  details depends on
somewhat on the absorption corrections.

Now we are in the position to evaluate the total cross section for
$\pi^+\pi^-$ scattering. The results for the Regge-pole
approximation, including small hard two-gluon component, are shown
in Fig.\ref{fig_tot}a by the dashed line.
To the extent that the absorption corrections are small and depend
only weakly on the diffraction slope, the Regge factorization
evaluation of $\sigma_{tot}^{\pi\pi}$ is parameter free. This can
be judged from the thick solid line in which we include in
addition the absorption corrections evaluated in the
double-scattering approximation to be discussed in more detail in
the next section \footnote{Adding absorption corrections to the
Regge-factorization estimates for $C_{i}$ is not entirely
consistent, here we only want to give an idea on the size of the
absorption effects.}. Our predictions well coincide with total
cross sections extracted in \cite{ZS84} from the absorption Regge
model analysis of $\pi N \to XN, X\Delta$ reactions. This
extraction of the pion exchange contribution is not entirely
parameter free and is subject to $\sim$ (10-20)\% uncertainties.
The low-energy extrapolation of this parameter-free Regge model
results joins smoothly with the low energy data on the
$\pi^+\pi^-$ total cross section in the resonance region
\cite{Robertson,Protopopescu}, in close analogy to to the
situation in $\pi N,NN, KN,\bar{K}N$ scattering, see the total
cross section plots in PDG \cite{PDG}.

In Fig.\ref{fig_tot}b we compare our predictions for same-sign
pion-pion scattering with the quasi-data from \cite{ZS84}.
While the opposite-sign pion-pion total cross section depends strongly
on energy, the same-sign pion-pion total cross section is almost
independent of energy.
In the spirit of duality, the near cancellation of contributions
from crossing-even and crossing-odd Regge exchanges
in the $\pi^+\pi^+,\pi^-\pi^-$ channels
is not accidental and is consistent with the absence of isotensor
$s$-channel resonances, in close analogy to the flatness of
the $pp$ total cross section.
Exactly the same effect can be seen in the quasi-data from
\cite{ZS84}.

For the sake of completeness we show in Fig.\ref{fig_retoim} also
the ratio of the real-to-imaginary parts of the scattering amplitude:
\begin{equation}
\rho = \frac{Re(A(t=0,W))}{Im(A(t=0,W))}\, .
\label{real_to_imag}
\end{equation}
In close analogy to the $p\bar{p},pp$ system, the destructive
interference of crossing-odd and crossing-even amplitudes in the
total cross section corresponds to the constructive interference
in the real part and vice versa. For this reason $\rho$ is
large in the $\pi^{+}\pi^+$ and $\pi^-\pi^-$ channels and
nearly negligible in the $\pi^{+}\pi^-$ channel.

\section{Absorption corrections and multiple soft and hard exchanges}

When going to the region of intermediate $|t|$ one has to include
absorption corrections, which model the Regge cuts due to multiple
pomeron and reggeon exchanges. The salient feature of these
multiple exchange amplitudes is that at large $|t|$ they decrease
slower than the Regge-pole amplitude. At high energies, when the
pomeron exchange dominates the leading absorption corrections are
known to have the sign opposite to single reggeon exchange. This
is the origin of diffractive dip in, e.g., proton-proton
scattering which is partly filled due to a smaller real part of
the scattering amplitude. At intermediate energies, we consider
here, the situation is somewhat more complicated. The absorption
corrections affect also substantially the diffraction slope and
its $t$-dependence. In this section we shall discuss such effects
for pion-pion scattering.

In the evaluation of absorption corrections one usually resorts to
the so-called eikonal approximation (see for instance \cite{T-M}
and references therein; a more recent discussion and references
can be found e.g. in \cite{Levin}). Here we restrict ourselves to
the dominant double-scattering corrections which read
\begin{equation}
A_{ij}^{(2)}(s,\vec{k}) =
\frac{i}{32 \pi^2 s} \int d^2 \vec{k}_1 d^2 \vec{k}_2 \;
\delta^2 (\vec{k} - \vec{k}_1 - \vec{k}_2) \;
A_i^{(1)}(s,\vec{k}_1) \; A_j^{(1)}(s,\vec{k}_2) \; .
\label{double_scattering}
\end{equation}
In general, the single scattering amplitudes $A_k^{(1)}$ in
(\ref{double_scattering}) are not restricted to soft reggeon
exchanges and hard two-gluon exchanges should be included too.
Consequently in the following we shall include the (soft $\otimes$
soft), (soft $\otimes$ hard)+(hard $\otimes$ soft) and (hard
$\otimes$ hard) double-scattering amplitudes. The
double-scattering involving hard mechanism is expected to be
small, at least at forward angles, compared to the leading (soft
$\otimes$ soft) absorption correction. In the (hard $\otimes$
hard) case the eikonal amplitude sums only a certain subset of
possible four-gluon exchange amplitudes, but numerically this
contribution is entirely negligible.

\begin{table}[t]

\caption{Global characteristics of elastic pion-pion
scattering for exponential ($B$=4 GeV$^{-2}$) and monopole
($B_{mon}$ = 1 GeV$^{-2}$) form factors.}

\begin{center}

\begin{tabular}{||c|c||c|c|c||c|c|c||}
\hline \hline \multicolumn{2}{||c||}{} &
\multicolumn{3}{c||}{opposite-sign pions} &
\multicolumn{3}{c||}{same-sign pions} \\
\hline reggeon &
 W (GeV) & $\sigma_{el}^{1}$& $\sigma_{el}$ &
           $ B_{eff}$  &
           $\sigma_{el}^{1}$ & $\sigma_{el}$  &
           $ B_{eff}$  \\
 f.f. &
 W (GeV) & (mb) & (mb) &
           (GeV$^{-2})$ &
            (mb) & (mb) &
           (GeV$^{-2})$ \\
\hline \hline exp.
     & 3 & 2.87  & 1.38  & 7.59 & 1.05  & 0.91 & 8.62
\\
     & 4 & 2.10 & 1.20  & 7.91 & 0.86  & 0.72  & 8.81
\\
     & 5 & 1.71  & 1.06  & 8.08 & 0.77 & 0.64  & 8.73 \\
\hline \hline
mon.
     & 3 & 3.13  & 1.38  & 7.77 & 1.13  & 0.96 & 8.63 \\
     & 4 & 2.28  & 1.20  & 8.07 & 0.93  & 0.76 & 8.85 \\
     & 5 & 1.84  & 1.08  & 8.21 & 0.85  & 0.68 & 8.79 \\
\hline \hline
\end{tabular}

\end{center}

\end{table}

First, let us focus  on salient features of double-scattering
contributions to the imaginary part of the pion-pion elastic
scattering amplitude shown  in Fig.\ref{fig_double_scattering}. In
this calculation $m_g$ = 0.75 GeV and the slope parameter B = 4
GeV$^{-2}$ is adjusted to have a reasonable slope of the forward
diffractive peak. We observe that at large $|t|$ the mixed (soft
$\otimes$ hard) + (hard $\otimes$ soft) terms are of a size
comparable to that of the (soft $\otimes$ soft) terms. Very small
(hard $\otimes$ hard) terms can at best contribute at the
diffraction dip, otherwise it is negligible small for all the
practical purposes. We observe a huge difference between the
opposite-sign and same-sign pion-pion scattering for the
(soft$\otimes$soft component, which is caused by a substantial
contribution from the secondary reggeon exchange. This difference
is further illustrated in Fig.\ref{fig_Im_A}, where we present for
the decomposition of the imaginary part of the scattering
amplitude for elastic $\pi^+\pi^-$ scattering (left panel) and the
same-sign pion scattering (right panel) at W = 4 GeV (solid lines)
into single- (dash-dotted line) and double-scattering
contributions (dashed line) terms. A destructive interference of
the single- and double-scattering contributions is much weaker for
the same-sign pions than for the opposite-sign pions. This
property is also seen in the table 2, where we show the total
elastic cross section calculated for several energies in the Regge
impulse approximation without, $\sigma_{el}^1$, and including,
$\sigma_{el}$, absorption corrections (we notice in passing that
the effect of the pQCD 2G-exchange on total elastic cross section
is marginal and does not exceed 10 per cent) . The two evaluations
do practically coincide for the same-sign pions, but for the
opposite-sign pions the effect of absorption on elastic cross
section is strong. That will lead to a pronounced difference for
the corresponding cross sections in the soft-hard interference
region of intermediate $|t|$.


In Fig.\ref{fig_1st_2nd} we show the decomposition of the
differential cross sections of elastic scattering (solid line)
into the contributions of single- , including both the sift and 2G
exchange contributions, and double-scattering terms, shown by the
dashed and dotted lines, respectively (the interference term is
not shown). One observes quite a different pattern of destructive
interference of single- and double-scattering for the
opposite-sign (left panel) and the same-sign (right panel)
pion-pion scattering. The diffractive dips for the opposite-sign
pion-pion scattering occur at values of $t$ at which the
contributions to differential cross section from single- (dashed
line) and double-scattering (dotted line) are about identical.
Whereas the origin of different diffractive structures for the
same-sign and opposite-sign pions is clear and the onset of pQCD
hard regime in the same-sign pion scattering is substantiated by
this analysis, one must conclude that model-dependence of the
double-scattering amplitude makes the large-$|t|$ results and the
dip positions for the opposite-sign case numerically unstable.
This point is corroborated by the energy dependence of
differential cross section shown in fig.\ref{fig_diff_energies}:
while for the same-sign pions (the right panel) it is very weak in
accordance to the small contribution from secondary reggeons  to
the imaginary part of the scattering amplitude, for the
opposite-sign $\pi^+\pi^-$ case (the left panel) the substantial
energy dependence of the secondary reggeon contribution entail
very strong energy dependence at large-$t$. Only at very large $t
\sim$ 4 GeV$^2$ the $\pi^+\pi^-$ indicates an onset of weak energy
dependence. This masking effect of subleading reggeon exchanges
persists up to rather high $W \sim$ 10-20 GeV.

The above shown results were for the exponential reggeon-pion
vertices with $B=4$ GeV$^-2$. The sensitivity of the forward
diffraction peak to the parameterization of the reggeon-pion
vertices is shown in fig.\ref{fig_eff_slope}, where we plot the
effective diffraction slope
$$
B_{eff} = -\frac{log(d\sigma/dt)}{dt}
$$
for the $\pi^+\pi^-$ scattering. These plots demonstrate clearly a
substantial enhancement of the diffraction slope by absorption
corrections: systematically $B_{eff}> B$, see also the results for
the diffraction slope shown in table 2. Notice the rise of
$B_{eff}$ with increasing $|t|$ for the exponential
parameterization, which is driven by destructive interference of
the single- and double-scattering amplitudes (the same trend is
obvious from the convex shape of the $\pi^+\pi^-$ differential
cross section in fig.~10). On the other hand, the diffraction
slope obtained with the monopole parameterization for the
reggeon-pion vertex decreases with rising $|t|$ in close semblance
to what has been observed experimentally in the proton-proton and
pion-proton scattering \cite{Schiz81,Burq83}. Still, at large
$|t|$ the monopole form factor gives a soft amplitude which does
not vanish much faster than pQCD  asymptotic predictions
\cite{elastic_pQCD} and, consequently, should not be applied
beyond the small $t$ region. This shows that there is no simple
one-parameter functional form which would be preferable for both
small and large $|t|$.
\section{Conclusions}

In the present analysis we have explored the onset of pQCD hard
mechanism for elastic pion-pion scattering in the region of
intermediate energies W = 3 - 5 GeV. The Glauber-Gribov-Landshoff
mechanism is shown to dominate hard two-gluon scattering at large
$|t|$. We have shown that while in the non-relativistic
approximation for the pion the GGL amplitude is free of the form
factor suppression at large $|t|$, this is not so in the
relativistic lightcone approach. Furthermore, the correlation
between the transverse and longitudinal motion of quarks inherent
to the lightcone treatment makes the GGL amplitude free of the
end-point contributions. Assuming dominance of soft physics and
Regge factorization at small $|t|$ we have predicted the total
cross section for pion-pion scattering consistent with
experimental values extracted in the literature.

We analysed the impact of large-$|t|$ tail of soft hadronic
scattering on the onset of pQCD hard mechanism. Within the
conventional Regge absorption  models, there emerges a rather
complex interplay of soft-hard interference. Specifically, pQCD
hard scattering is found to dominate elastic scattering of the
same sign pions at $|t| \gsim $3 GeV$^2$. However, in the case of
the opposite-sign pions the destructive soft-hard interference
remains strong up to at least $|t| \gsim $4 GeV$^2$.

The effects discussed here are important in the context of a
recently reported deficit of pQCD result at large-angle scattering
in $\gamma \gamma \rightarrow \pi^+\pi^-$ as compared to the
experimental data measured at electron-positron colliders. The
discussion of the latter goes beyond the scope of the present
analysis and will be presented elsewhere.

\vskip 1cm

{\bf Acknowledgment}
One of us (A.S.) is indebted to Katarzyna Grzelak and Carsten Vogt
for interesting discussion which initiated this work.
This work was partially supported by
the German-Polish DLR exchange program, grant number POL-028-98.


\newpage


\vskip 0.2cm


\begin{figure}

\hspace{1cm}
\epsfxsize 8cm
\epsffile{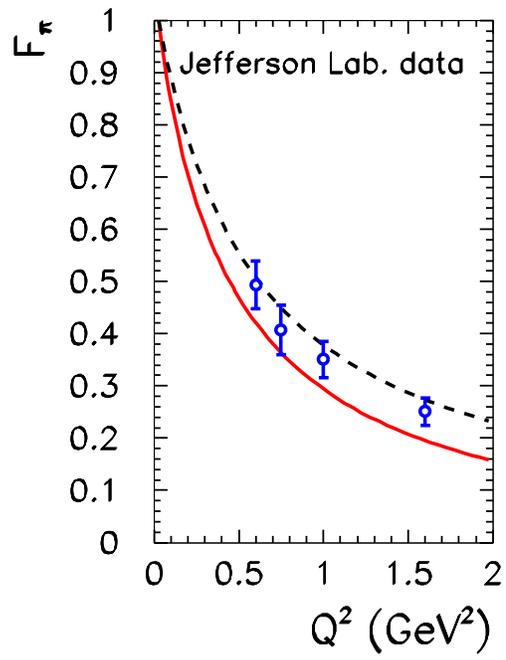}

\caption{\it
Charged pion electromagnetic form factor.
The experimental data are from \cite{CEBAFpionFF}.
The solid line is the result of calculation based on
Eq.(\ref{one_body_formfactor})
with the light-cone wave function.
The dashed line is the monopole parametrization with
the $\rho$-meson mass.
}
\label{fig_EM_ff}
\end{figure}


\vskip 0.2cm


\begin{figure}


\epsfxsize 6.0cm
\epsffile{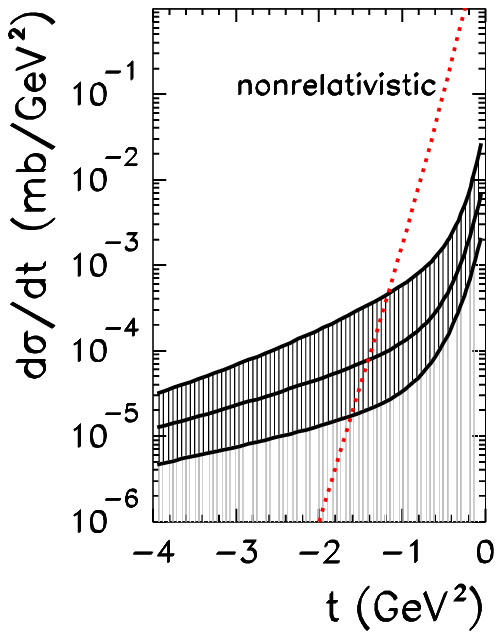}
\hspace{-1cm}
\epsfxsize 6.0cm
\epsffile{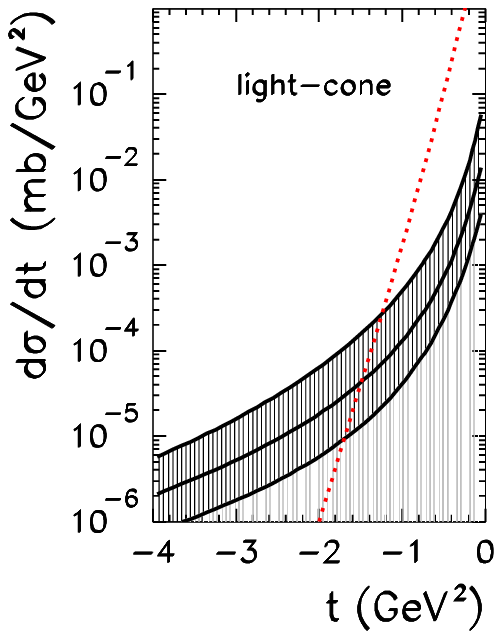}

\caption{\it Angular distribution of elastic pion-pion scattering
for the two-gluon exchange model with different values of the
infrared regularization parameter $m_g$ = 0.5, 0.75, 1.0 GeV. The
dashed line shows the soft $\pi^+\pi^-$ elastic scattering
evaluated in the Regge-pole approximation with B$_{eff}$ = 6
GeV$^{-2}$. } \label{fig_mg}
\end{figure}


\vskip 0.2cm

\begin{figure}

\epsfxsize 6.0cm
\epsffile{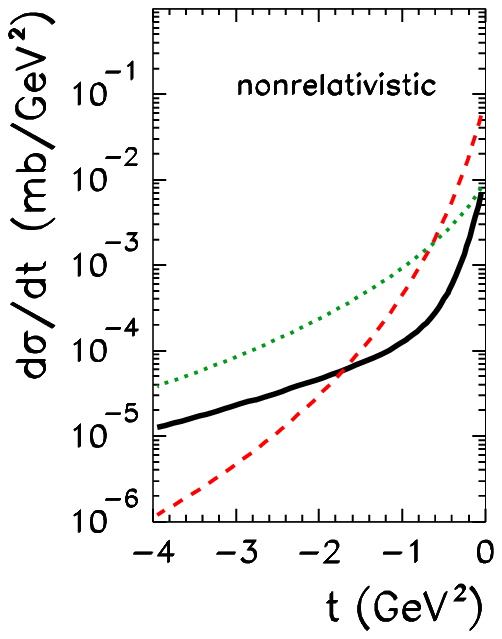}
\hspace{-1cm}
\epsfxsize 6.0cm
\epsffile{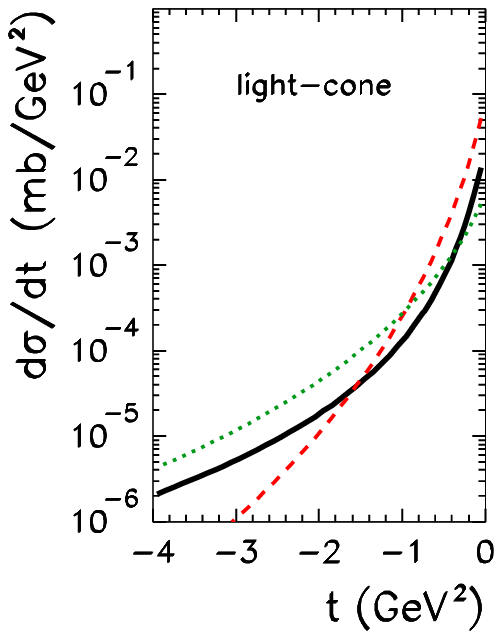}

\caption{\it
The emergence of GGL dominance in
elastic pion-pion scattering for the
two-gluon exchange model for $m_g$ = 0.75 GeV.
The dashed line is for pure IA contributions (a) and (b)
for the impact factor in Fig.\ref{fig_gluon},
whereas the dotted line corresponds to the pure GGL
terms (c) and (d) for the impact factor in Fig.\ref{fig_gluon}.
The thick solid line corresponds to the full result with
all terms for the impact factor.
}
\label{fig_pQCD_deco}
\end{figure}


\vskip 0.2cm


\begin{figure}


\epsfxsize 6.0cm
\epsffile{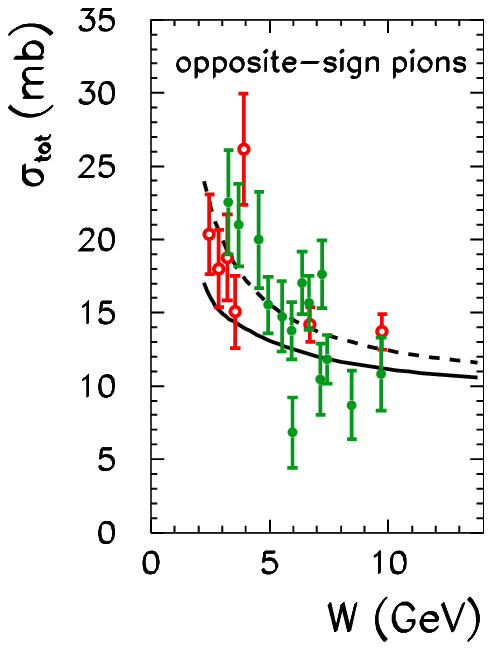}
\hspace{-1cm}
\epsfxsize 6.0cm
\epsffile{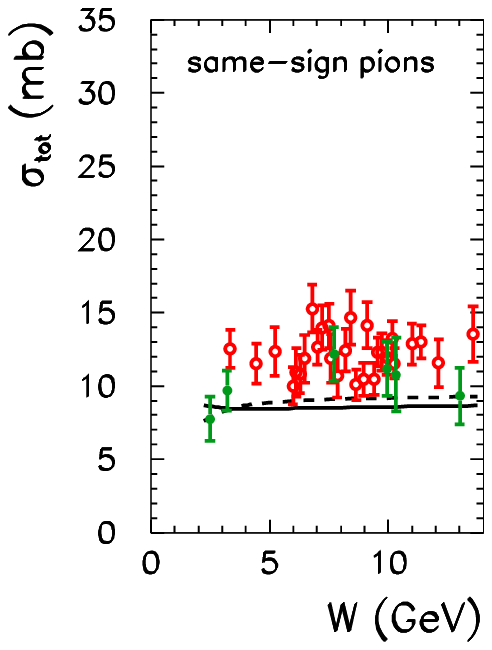}

\caption{\it
Total cross section for $\pi^+ \pi^-$ (left panel)
and $\pi^+\pi^+$ or $\pi^-\pi^-$ (right panel) scattering
as a function of center-of-mass energy $W$.
The experimental data are from \cite{ZS84}.
The experimental data for $\pi^+ \pi^-$ scattering (left panel)
were extracted from $\pi^+ p \rightarrow X \Delta^{++}$ (open circles)
and from $\pi^+ n \rightarrow Xp$ (full circles).
The experimental data for $\pi^- \pi^-$ scattering (right panel)
were extracted from $\pi^- p \rightarrow X \Delta^{++}$ (open circles)
and from $\pi^- n \rightarrow Xp$ (full circles).
The single pomeron and subleading reggeon exchanges
are given by the dashed lines. The solid line is obtained from the
dashed line after including the absorption corrections
to be discussed in the next section.
}
\label{fig_tot}
\end{figure}


\vskip 0.2cm


\begin{figure}

\epsfxsize 6.0cm
\epsffile{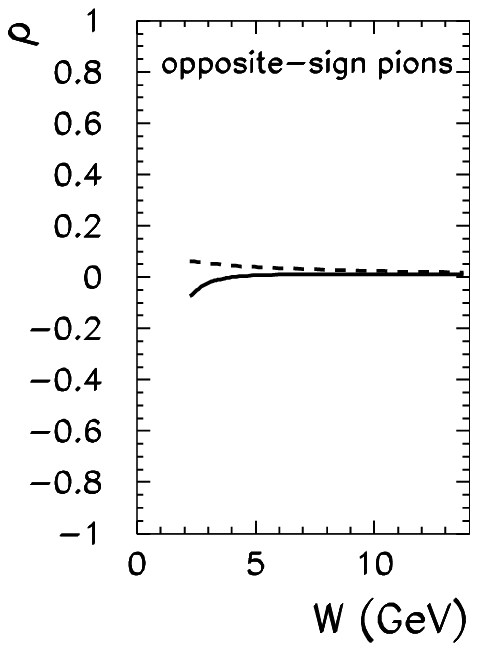}
\hspace{-1cm}
\epsfxsize 6.0cm
\epsffile{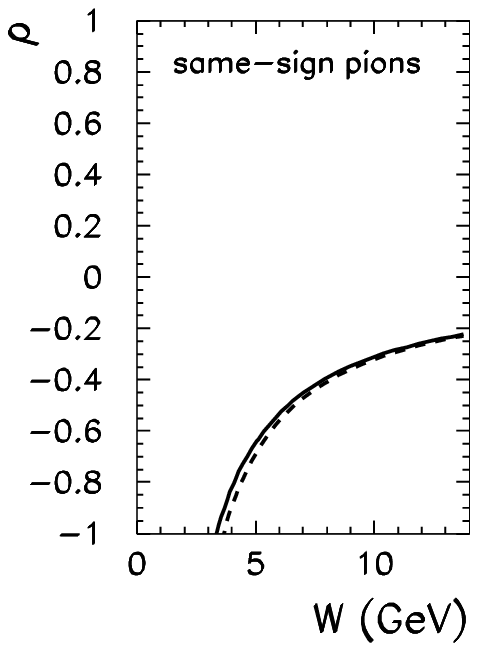}

\caption{\it
The ratio of the real to imaginary part of the forward
elastic scattering amplitude ($\rho$) as a function of
center-of-mass energy $W$. The meaning of the curves is
the same as in the previous figure.
}
\label{fig_retoim}
\end{figure}


\vskip 0.2cm


\begin{figure}

\epsfxsize 6cm
\epsffile{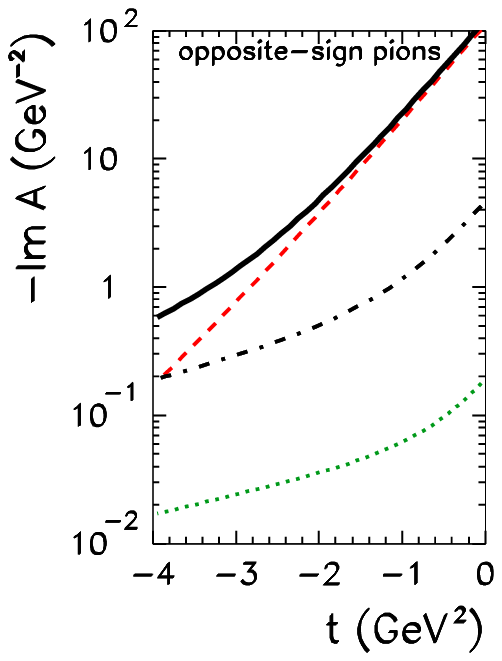}
\hspace{-1cm}
\epsfxsize 6cm
\epsffile{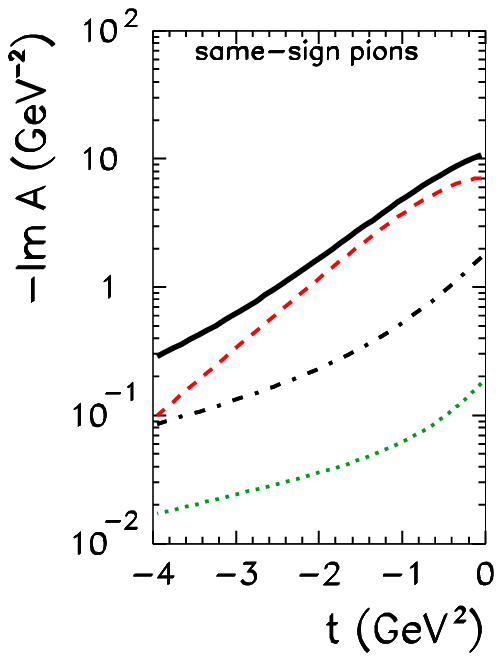}

\caption{\it
The imaginary part of the amplitude of isolated different
double-scattering terms for elastic $\pi \pi$ scattering:
(soft $\otimes$ soft) - dashed line,
(soft $\otimes$ hard) or (hard $\otimes$ soft) - dash-dotted line
and (hard $\otimes$ hard) - dotted line for W = 4 GeV
for the opposite-sign pions (left panel)
and for the same-sign pions (right-pions).
The sum of all contributions is given by the thick solid line.
}
\label{fig_double_scattering}
\end{figure}


\vskip 0.2cm


\begin{figure}

\epsfxsize 6cm
\epsffile{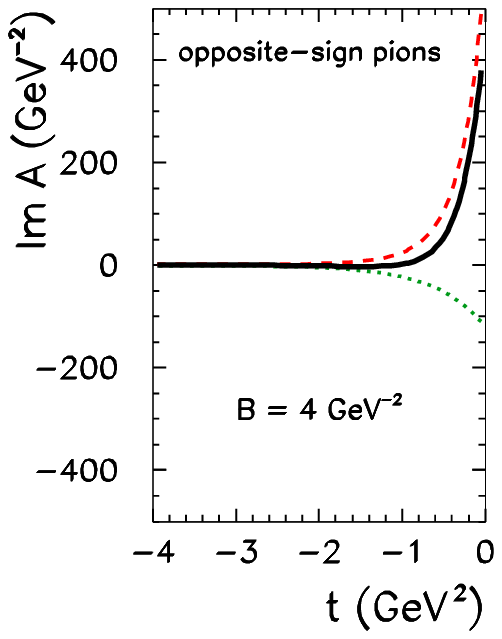}
\hspace{-1cm}
\epsfxsize 6cm
\epsffile{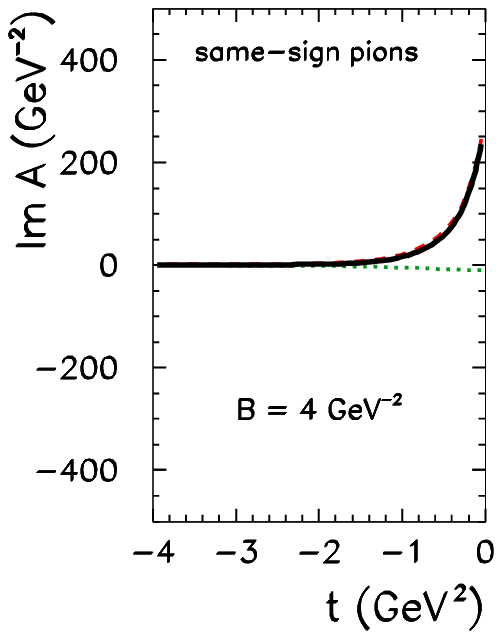}

\caption{\it
Imaginary part of the single-scattering (dashed) and
double-scattering (dotted) terms. The resultaing imaginary part
of the ful amplitude is shown by the solid line.
}
\label{fig_Im_A}
\end{figure}


\vskip 0.2cm


\begin{figure}

\epsfxsize 6cm
\epsffile{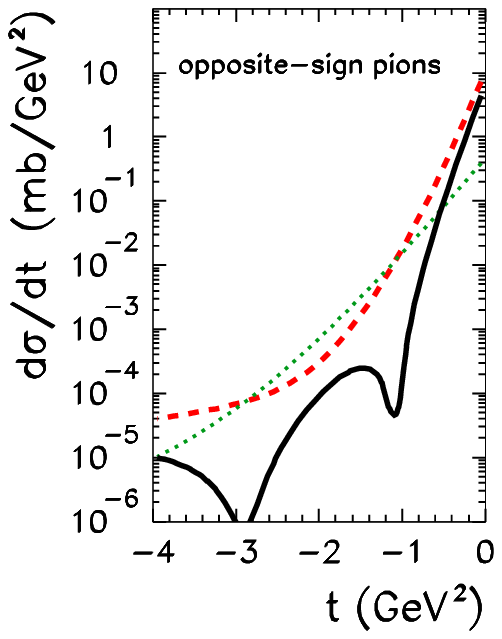}
\hspace{-1cm}
\epsfxsize 6cm
\epsffile{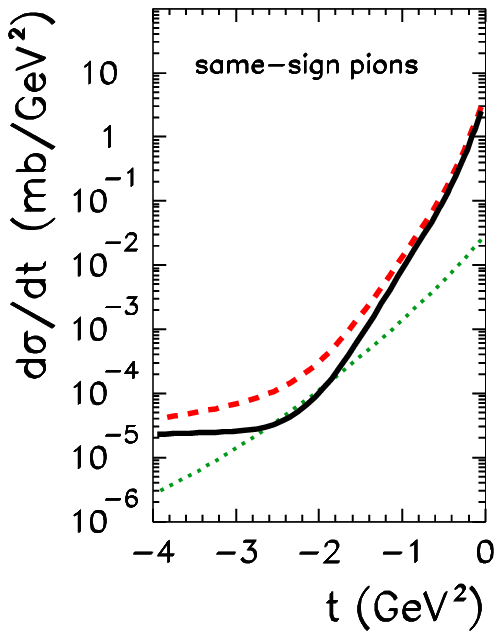}

\caption{\it
The effect of the absorption corrections on the $t$-dependence
of the elastic $\pi \pi$ cross sections for
opossite-sign pions (left panel) and same-sign pions (right panel)
for W = 4 GeV.
In this calculation the slope parameter B = 4 GeV$^{-2}$.
The cross section for single-exchange is shown by the dashed
line, while the cross section which includes double-scattering
effect by the solid line.
For the discussion in the text by the dotted line we show
the cross section calculated from the double-scattering
amplitude alone.}
\label{fig_1st_2nd}
\end{figure}


\vskip 0.2cm


\begin{figure}

\epsfxsize 6cm
\epsffile{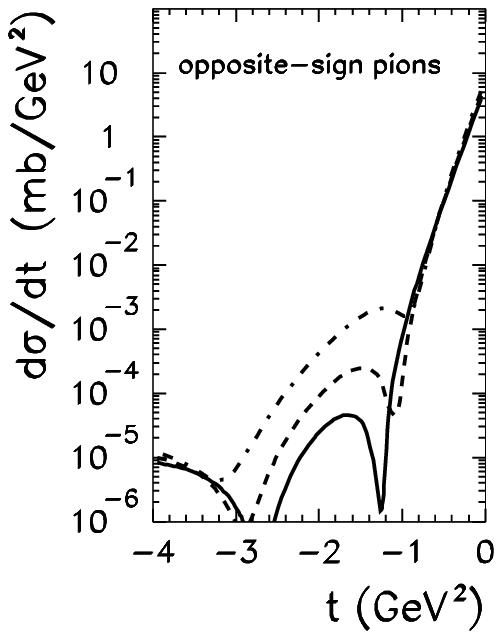}
\hspace{-1cm}
\epsfxsize 6cm
\epsffile{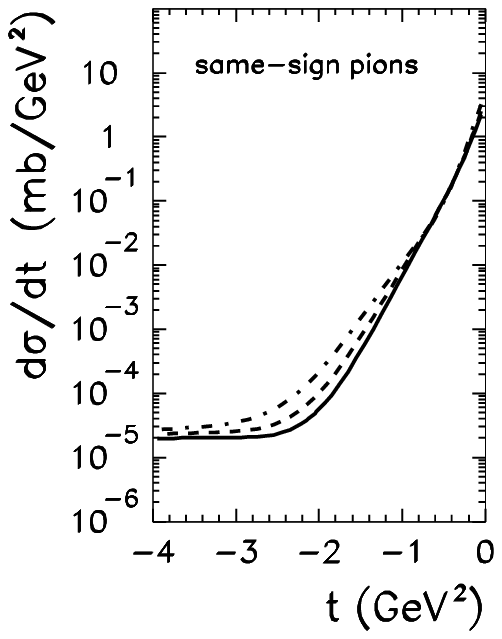}

\caption{\it
The $t$-dependence of the pion-pion elastic cross section
with the inclusion of all single- and double-exchange contributions
for different center of mass energies
 3 (dash-dotted),
4 (dashed), 5 (solid) GeV.
In this calculation: B = 4 GeV$^{-2}$, $m_g$ = 0.75 GeV.
}
\label{fig_diff_energies}
\end{figure}


\vskip 0.2cm


\begin{figure}

\epsfxsize 6.0cm
\epsffile{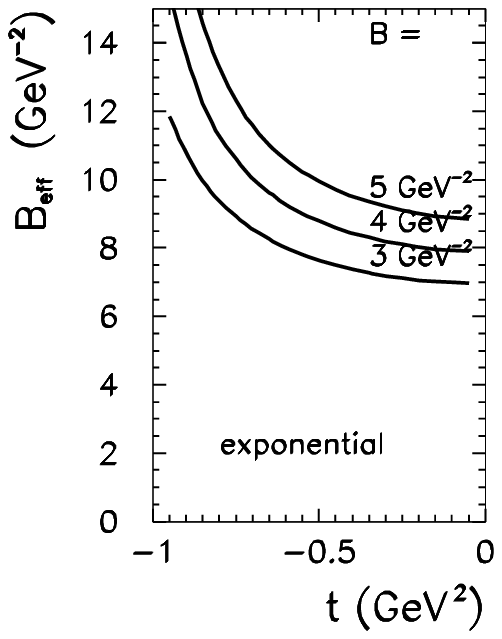}
\hspace{-1cm}
\epsfxsize 6.0cm
\epsffile{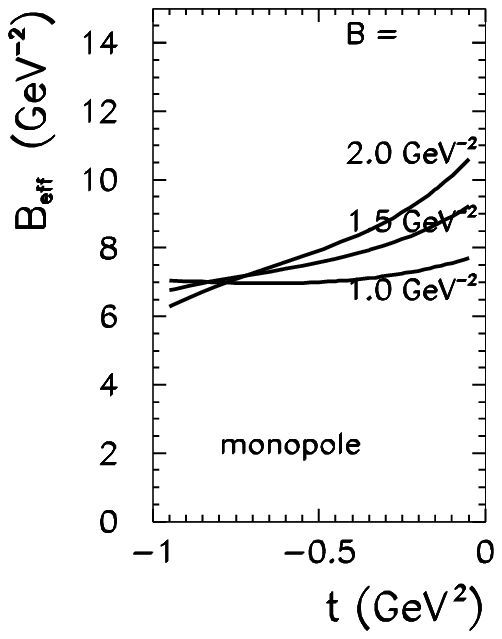}

\caption{\it
The effective slope parameter for elastic $\pi^+\pi^-$
scattering at W = 4 GeV as a function of the Mendelstam
variable $t$
for exponential ($B$ = 2, 3, 5 GeV$^{-2}$) and monopole
($B_{mon}$ = 1, 1.5, 2 GeV$^{-2}$) form factors.
}
\label{fig_eff_slope}
\end{figure}


\newpage




\begin{thebibliography}{99}


\bibitem{Weinberg}
S. Weinberg,
Phys. Rev. Lett. {\bf 18} (1967) 188.

\bibitem{ChPT}
S. Weinberg,
Physica {\bf A96} (1979) 327;\\
J. Gasser and H. Leutwyler,
Ann. Phys. NY {\bf 158} (1984) 142.

\bibitem{effective_field_theories}
G. Ecker, J. Gasser, A. Pich and E. de Rafael,
Nucl. Phys. {\bf B321} (1989) 311;\\
V. Bernard, N. Kaiser and U.G. Meissner,
Nucl. Phys. {\bf B364} (1991) 283;\\
J.A. Oller and E. Oset,
Phys. Rev. {\bf D60} (1999) 074023

\bibitem{unitarization}
J.A. Oller, E. Oset and A. Ramos,
Prog. Part. Nucl. Phys. {\bf 45} (2000) 157.

\bibitem{meson_exchange_models}
G. Janssen, B.C. Pearce, K. Holinde and J. Speth,
Phys. Rev. {\bf D52} (1995) 2690.\\
O. Krehl, R. Rapp and J. Speth,
Phys. Lett. {\bf B390} (1997) 23.

\bibitem{Brookhaven}
J. Gunter et al. (E852 Collaboration),
hep-ex/0001038.

\bibitem{ZS84}
B.G. Zakharov and V.N. Sergeev,
Sov. J. Nucl. Phys. {\bf 39} (1984) 448.


\bibitem{Glauber}
R.J. Glauber, Lectures in theoretical physics, vol.1, eds.
W. Brittain and L.G. Dunham (Interscience, New York, 1959);\\
R.J. Glauber and G. Matthiae, Nucl. Phys. {\bf B21} (1970) 135.

\bibitem{Gribov}
V.N. Gribov, Sov. Phys. JETP {\bf 29} (1969) 483; Zh. Eksp. Teor. Fiz.
{\bf 56} (1969) 892.

\bibitem{Landshoff}
P.V. Landshoff,
Phys. Rev. {\bf D10} (1974) 1024.

\bibitem{gamma_gamma}
H. Aihara et al. (TPC/Two-Gamma Collaboration),
Phys. Rev. Lett. {\bf 57} (1986) 404;\\
J. Dominick et al. (CLEO Collaboration),
Phys. Rev. {\bf D50} (1994) 3027.

\bibitem{pQCD}
S.J. Brodsky and G.P. Lepage,
Phys. Rev. {\bf D24} (1981) 1808;\\
B. Nizic,
Phys. Rev. {\bf D35} (1987) 80;\\
Ch.R. Ji and F. Amiri,
Phys. Rev. {\bf D3764} (1990) 3764;\\
C. Vogt, hep-ph/9911253, hep-ph/0010040.

\bibitem{Grzelak}
K. Grzelak, private communication.


\bibitem{2Gluon}
J.F. Gunion and D.E. Soper,
Phys. Rev. {\bf D15} (1977) 2617;
E.M. Levin and M.G. Ryskin,
Sov. J. Nucl. Phys. {\bf 34} (1981) 619.

\bibitem{QED}
V.N. Gribov, L.N. Lipatov and G.V. Frolov,
Sov. J. Nucl. Phys. {\bf 12} (1971) 543; \\
L.N. Lipatov and G.V. Frolov, Sov. J. Nucl. Phys. {\bf 13} (1971)
333.


\bibitem{Schwiete}
G. Schwiete, master thesis, Forschungszentrum J\"ulich, November
2000.

\bibitem{NSS01}
N.N. Nikolaev, W. Sch\"afer and G. Schwiete,
Phys. Rev. {bf D63} (2001) 014020.

\bibitem{Jaus}
W. Jaus,
Phys. Rev. {\bf D44} (1991) 2851.

\bibitem{Botts}
J. Botts, Phys. Rev. {\bf D44} (1991) 2768; J. Botts and G.
Sterman, Nucl. Phys. {\bf B325} (1989) 62; J. Bolz, R. Jakob, P.
Kroll, M. Bergmann and N.G. Stefanis, Z. Phys. {\bf C66} (1995)
267.

\bibitem{PDG}
Particle Data Group, C. Caso et al.,
Eur. Phys. J. {\bf C3} (1998) 209.


\bibitem{CEBAFpionFF}
J. Volmer et al., Phys. Rev. Lett. {\bf 86} (2001) 1713


\bibitem{NZZ}
N.N. Nikolaev and B.G. Zakharov, Phys. Lett. {\bf B327} (1994) 149;
N.N. Nikolaev, B.G. Zakharov and V.R. Zoller, JETP Lett. {\bf 66}
(1997) 138; Pisma Zh.Eksp.Teor.Fiz. {\bf 66} (1997) 134.

\bibitem{Meggiolaro}
E. Meggiolaro,
Phys. Lett. {\bf B451} (1999) 414.

\bibitem{Field}
J.H. Field, hep-ph/0101158.

\bibitem{Regge}
A.C. Irving and R.P. Worden,
Phys. Rep. {\bf 34} (1977) 117; \\
A.B. Kaidalov,
Phys. Rep. {\bf 50} (1979) 157.

\bibitem{Collins}
P.D.B. Collins, An Introduction to Regge Theory and High Energy Physics,
Cambridge University Press, Cambridge 1977.


\bibitem{DL92}
A. Donnachie and P.V. Landshoff,
Phys. Lett. {\bf B296} (1992) 227.

\bibitem{NSZpion}
N.N. Nikolaev, J. Speth and V.R. Zoller, Phys. Lett. {\bf B473} (2000) 157.

\bibitem{QCD_nonperturbative}
H.G. Dosch, E. Ferreira and A. Kr\"amer,
Phys. Rev. {\bf D50} (1994) 1992.

\bibitem{Robertson}
W.J. Robertson, W.D. Walker and L. Davies,
Duke University preprint, Durham, N.C. (May 1972).

\bibitem{Protopopescu}
S.D. Protopopescu et al.,
Phys. Rev. {\bf D7} (1973) 1279.


\bibitem{T-M}
K.A. Ter-Martirosyan,
Sov. J. Nucl. Phys. {\bf 10} (1970) 600.

\bibitem{Levin}
E. Gotsman, E. Levin and  U. Maor, Phys. Lett. {\bf B452} (1999) 387.

\bibitem{Schiz81}
A. Schiz et al., Phys. Rev. {\bf D 24} (1981) 26.

\bibitem{Burq83}
J.P. Burq et al., Nucl. Phys. {B217} (1983) 285.

\bibitem{elastic_pQCD}
S.J. Brodsky and G. Farrar, Phys. Rev. Lett. {\bf 31}
(1973) 1153;\\
S.J. Brodsky and G. Farrar, Phys. Rev. {\bf D11} (1975) 1309.


\end{thebibliography}
\end{document}